\documentclass[aps,prb,twocolumn,showpacks]{revtex4}
\usepackage{graphicx}  
\usepackage{dcolumn}   
\usepackage{bm}        
\usepackage{amssymb}   
\usepackage{braket}
\usepackage{color}
\usepackage{soul}

\begin{document}

\widetext

\setlength{\textwidth}{175.0mm}


\title{Design of nonlinear optical response of multipole-type excitons
\\ by film thickness and incident pulse width} 
\date{\today}

\author{Takashi Kinoshita}
\author{Hajime Ishihara}
\affiliation{Department of Physics and Electronics, Osaka Prefecture University, Sakai, Osaka 599-8531, Japan}

\begin{abstract}
We theoretically investigate the nonlinear optical pulse responses of excitons in a thin film where the excitonic center-of-mass motion is confined. 
A large interaction volume between excitons and radiation yields particular coupled states with radiative decay times reaching several femto-seconds.
By considering two polarization directions of light, we reveal that these fast-decay modes dominantly survive in an optical Kerr spectra even under a massive nonradiative damping $\Gamma=30$ meV.
The results clearly show that there is an optimal combination of the incident pulse width and the film thickness for maximizing the integrated intensity of nonlinear signals. 
\end{abstract}

\maketitle



\section{INTRODUCTION}
  An attractive feature of nanostructures is their great accessibility to single quantum states due to their apparent quantization, which allows us to develop unconventional photo-functions owing to the flexible controllability of the light-matter interaction \cite{Meier,Gavrilenko}. 
However, the light-matter interaction of a single quantum state is essentially weak because of the localization of its wavefunction. 
An effective method for overcoming the small reaction cross-section of a single quantum state is utilizing auxiliary systems such as microcavities or optical antennas made with metallic structures, where the extremely localized photonic modes realize a high probability of excitation of the single quantum states by a few photons \cite{Kasprzak,Xiao,Osaka}. 
An other solution involves the realization of spatially extended quantum states. 
A large coherent volume due to the collective dipole motions leads to an enhanced oscillator strength \cite{Hanamura,Takagahara}.
In particular, the coherence length of excitonic center-of-mass (CM) motions reaches several hundreds of nano meters in the one-dimensional confined system.
This situation yields a large exciton-radiation coupling because the  multipole-type excitons with the CM quantum number $\lambda\geq2$ can match their spatial phase with the radiation wave\cite{Ishihara,Syouji,Ichimiya1}.
In contrast to the optical responses based on the conventional long-wavelength approximation (LWA) of light, this type of interaction exhibits remarkable optical effects such as a large radiative shift leading to the interchange of the quantized levels \cite{Syouji} and ultrafast radiative decay \cite{Ichimiya1}.    
Such exciton--radiation-coupling beyond the LWA strongly modifies the optical spectra relative to those expected from bare excitonic systems, and the interpretation of the spectral shape becomes complex. 
Although numerous studies have been performed on the excitonic properties in the thin film geometry \cite{DelSole2,Agarwal,CK,CI,Tredicucci,Beloussov,Foy}, efforts to exploit the potential of light-matter coupling beyond the LWA regime have only just begun, and the unresolved physics and potential applications of excitons in promising materials offer scope for further research. 

Nanostructures of ZnO have attracted the attention of researchers because of their potential applications in optoelectronic devices operable even at room temperature (RT), such as light-emitting diodes \cite{Min,Li}, ultraviolet photovoltaics \cite{Cole}, and exciton polariton lasing \cite{Orosz,Feng} utilizing the wide band gap and large exciton binding energy. 
Although the formation of an exciton--radiation-coupled system is a key to the exploitation of such applications, even the fundamental structures of the coupled modes were previously unclear. 
In recent years, however, we revealed that the radiation-induced coupling between A and B excitons in thin film structures enhances the radiative decay rate of particular coupled states \cite{Kinoshita}.
This result motivates us to exploit coherent nonlinear photo-functions such as ultrafast optical switching at RT because the enhanced decay rates can be expected to exceed the thermal dephasing at RT \cite{Ichimiya2} (typically several tens of or several hundreds of femto-seconds).

Herein, by considering two polarization directions of light, we expand the nonlocal response theory of multicomponent excitons \cite{Kinoshita} and investigate the optical Kerr response (OKR), which is known as a typical third-order nonlinear optical effect.
Then, we demonstrate that a particular A-B-coupled state in ZnO with an enhanced radiative width of over 50 meV dominantly survives in the OKR signals, even under a massive nonradiative damping $\Gamma=30$ meV.
The incident pulse width and the film-thickness dependence of the integrated intensity of nonlinear signals clearly show that there is an optimal combination between these two parameters for enhancing the optical nonlinearity.
The results indicate the importance of an integrated design of nanostructures and an input optical pulse for maximizing the veiled potential of single quantum states for ultrafast nonlinear optics. 
The presented demonstrations exhibit a striking contrast to the conventional understandings of excitons, where the coherent nonlinear response is considered to be negligible at RT because the radiative decay rate of excitons never exceeds the thermal dephasing.  

  The rest of this article is organized as follows:
Section \ref{2} outlines the theory of nonlocal optical response for deriving the radiative decay time of the exciton--radiation-coupled system.
Section \ref{3} demonstrates the optical Kerr spectra considering two kinds of temperature regions: cryogenic temperature (CT) and RT.
In Sec. \ref{4}, we investigate the pulse width and film thickness dependences of nonlinear signals to determine the optimal pulse width for each film thickness.
The results and discussions in this article are summarized in Sec. \ref{5}.

\section{Radiative decay time\label{2}}
  Beyond the LWA regime, the interplay between the spatial structures of the radiation and excitonic wavefunction is activated.
Recently, we constructed a theoretical framework for the nonlocal optical response of multicomponent excitons in a thin--film structure with special attention to their self-consistency \cite{Kinoshita}. 
In this section, we review the formalism of the linear response to obtain a full understanding of exciton--radiation coupled modes and the origin of fast radiative decay. 
As a model system, we considered a thin-film structure with a thickness of $d$ along the $z$ axis.
The thickness is assumed to be far greater than the effective Bohr radius of an exciton; thus, the excitonic relative motion can be treated in the same way as that in the bulk system.
In this condition, only the CM motion of excitons is confined in the sample.
We neglect the confinement effect of the relative motions of electron--hole pair, which dominantly contributes to the energy structure of excitons in the size region where the thickness reaches the effective Bohr radius \cite{Kayanuma} (～1.8 nm for ZnO \cite{Klingshirn}). 
According to the standard effective-mass approximation, the eigenenergy of the bare exciton is given as $E_{\sigma\lambda}=E_\sigma+(\hbar^2k^2_{\sigma\lambda})/(2M_\sigma)$, where $\sigma$ is an index for labeling multiple exciton bands (thus, $\sigma$ corresponds to A or B in the case of ZnO), $E_\sigma$ is the energy of a transverse exciton in the bulk limit, and $M_\sigma$ is the effective mass of an exciton. 
In a thin sample, the distortion of wavefunctions near the surface generally affects the energy structures of excitons.
We therefore applied a microscopic transition layer model \cite{DelSole,Ishihara2} as the excitonic CM wavefunction $g_{\sigma\lambda}(z)$.
In this model, the quantization condition is given as 
$
k_{\sigma\lambda}d-2\tan^{-1}{k_{\sigma\lambda}/P_\sigma}=\lambda\pi
$
($\lambda=$ $1,2,\cdots$), where $P_\sigma$ is a decay constant of evanescent waves with a value on the order of the inverse of the effective Bohr radius, indicating the distortion length.
In this paper, we fix these values as the effective Bohr radius ($1/P_A=1/P_B=1.8$ nm) because the optical signal is not sensitive to a change in $P_\sigma$ for a thickness beyond the LWA regime \cite{Kinoshita}, although for thin samples in the LWA regime, $P_\sigma$ is one of the important parameters for accurate analysis of the CM quantization as demonstrated in Ref. \cite{Yoshimoto}.

According to the linear-response theory \cite{Kubo}, the $j$th-order polarization can be obtained from the perturbation expansion method of the density matrix.  
The first-order polarization in the site representation ${\cal P}^{(1)}(z,\omega)$ can be written in nonlocal form as \cite{Kinoshita}

\begin{eqnarray}
{\cal P}^{(1)}(z,\omega)
=
\int\chi(z,z',\omega){\cal E}(z',\omega)dz'.
\end{eqnarray}
In this expression, a resonant term of the nonlocal susceptibility is written as 

\begin{eqnarray}
\chi(z,z',\omega)
=
\sum_{\sigma}\sum_{\lambda}\frac{p_{\sigma\lambda}(z)p^\ast_{\sigma\lambda}(z')}{E_{\sigma\lambda}-\hbar\omega-i\Gamma_\sigma},
\label{chi}
\end{eqnarray}
where $\Gamma_\sigma$ is a nonradiative damping constant and $p_{\sigma\lambda}(z)=\mu_\sigma g_{\sigma\lambda}(z)$.
In our definition, $\mu_\sigma$ has the dimension of a dipole moment per one-half power of volume.
This value is obtained from multiple longitudinal-transverse splitting energies \cite{Kinoshita}.

${\cal P}(z)$ should be determined self-consistently with the electromagnetic field in the Maxwell's equation.
Assuming normal incidence for simplicity, the Maxwell electric field ${\cal E}(z,\omega)$ in integral form is written as

\begin{eqnarray}
{\cal E}(z,\omega)
&=&
{\cal E}^{(0)}(z,\omega)
\nonumber 
\\
&+&
4\pi q^2\int dz'{\cal G}(z,z',\omega){\cal P}(z',\omega),
\label{maxwell}
\end{eqnarray}
where $E^{(0)}(z,\omega)$ is the background electric field, and $G(z,z',\omega)$ is the retarded Green's function for a thin-film structure \cite{Chew}.
The eigenmodes of an exciton--radiation-coupled system are obtained from

\begin{eqnarray}
{\rm det}|(E_{\sigma'\lambda'}-\hbar\omega-i\Gamma_{\sigma'})\delta_{\sigma'\sigma}\delta_{\lambda'\lambda}+A_{\sigma\sigma'\lambda\lambda'}(\omega)|=0,
\label{self-consistent}
\end{eqnarray}
where $A_{\sigma'\sigma\lambda'\lambda}(\omega)$ describes the radiative correction from the bare exciton state written as 

\begin{eqnarray}
A_{\sigma'\sigma\lambda'\lambda}(\omega)
&
\nonumber
\\
&\hspace{-10mm}=-4\pi q^2\int\int dzdz'p^\ast_{\sigma'\lambda'}(z){\cal G}(z,z',\omega)p_{\sigma\lambda}(z'),
\end{eqnarray}
which indicates the radiative coupling between the $\lambda$th $\sigma$-band exciton and $\lambda'$th $\sigma'$-band exciton via radiation.
This term also includes the radiation-induced coupling between different band excitons (A and B excitons for ZnO) when $\sigma'\neq\sigma$.

The real part Re[$\hbar\omega_{\xi}$] gives the eigenenergy including the radiative shift from the bare exciton energy, and the imaginary part $-$Im[$\hbar\omega_{\xi}$] gives the radiative width, where $\xi$ is an index of the quantized exciton--radiation-coupled states.
Considering the exponential decay of signals, we defined the radiative decay time $\tau_\xi$ as

\begin{eqnarray}
\tau_{\xi}=\frac{1}{-2{\rm Im}[\omega_{\xi}]}.
\end{eqnarray}

\section{Optical Kerr response \label{3}}
\begin{figure}[h]
\begin{center}
\includegraphics[width=85mm]{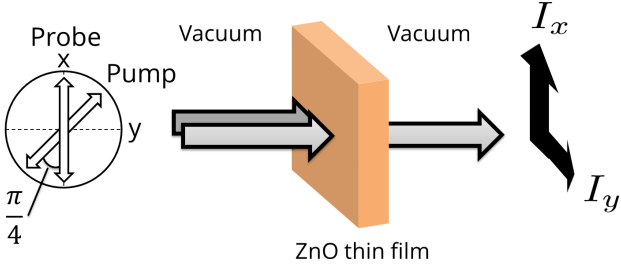}
\caption{Calculation model for optical Kerr response}
\label{setup2}
\end{center}
\end{figure}

First, we investigate the nonlinear optical spectra of A and B excitons in a ZnO thin film, focusing on the OKR. 
Figure \ref{setup2} shows a calculation model for the OKR. 
The polarization angle of the pump light is rotated by $\pi/4$ to that of the X-polarized probe light. 
The Y-polarized probe light contains a pure nonlinear component without the background electric field, i.e., ${\cal E}^{(0)}_y(z,\omega)=0$.
On the other hand, the X-polarized probe light contains both nonlinear and linear components, including the background electric field.
To calculate the output OKR signals, we expand the formalism of degenerate four-wave mixing (DFWM) in Ref. \cite{Kinoshita} with consideration of two polarization directions of light.
In the present demonstration, we focus on the dominant contribution, i.e., the effects of the one-exciton resonance, while avoiding non-essential issues of two-exciton contributions \cite{Kinoshita}.
Elaborate analysis considering the free two-exciton states through the cancellation effect \cite{Ishihara4} is necessary for evaluating the absolute values of OKR signals, although we do not consider it in the present study.

Considering the first- and third-order polarizations, 
the total electric field of this configuration can be written as

\begin{eqnarray}
{\cal E}_{x(y)}(z,\omega)&=&{\cal E}^{(0)}_{x(y)}(z,\omega) \nonumber 
\\
&&\hspace{-15mm}+\sum_\sigma\sum_\nu\{X^{x(y)}_{\sigma\nu}(\omega)+U^{x(y)}_{\sigma\nu}(\omega)\}
B_{\sigma\nu}(z,\omega),
\label{okr}  
\end{eqnarray}
where $B_{\sigma\nu}(z,\omega)$ is defined as

\begin{eqnarray}
B_{\sigma\nu}(z,\omega)
=
4\pi q^2\int dz'{\cal G}(z,z',\omega)p_{\sigma\nu}(z').
\end{eqnarray}
In Eq. (\ref{okr}), $X^{x(y)}_{\sigma\lambda}(\omega)$ and $U^{x(y)}_{\sigma\nu}(\omega)$ are written as

\begin{eqnarray}
X^{x(y)}_{\sigma\lambda}(\omega)
=
\frac{1}{E_{\sigma\lambda}-\hbar\omega-i\Gamma_\sigma}\int \,p^\ast_{\sigma\lambda}(z){\cal E}_{x(y)}(z,\omega)\,dz,
\end{eqnarray}
and

\begin{eqnarray}
&
&
\hspace{-5mm}
U^{x(y)}_{\sigma\nu}(\omega)
=
\sum_\lambda\int\int d\omega_1d\omega_2\bar{X}_{\sigma\nu\lambda}(\omega, \omega_1, \omega_2) \nonumber 
\\
&&
\times H^{pump_{x(y)}}_{\sigma\nu}(\omega_1)
H^{\ast pump_x}_{\sigma\lambda}((\omega_1+\omega_2)-\omega)
H^{probe_x}_{\sigma\lambda}(\omega_2), \nonumber
\\
\label{U}
\end{eqnarray}
where $H_{\sigma\nu}(\omega)=\int p^\ast_{\sigma\nu}(z){\cal E}(z,\omega)\,dz$ should be determined self-consistently by solving the third-order Maxwell's equation. 
However, if we assume that an electric field originating from the third-order polarization is far weaker than that originated from the linear polarization, this value corresponds well to the solution of the linear response calculation.
$\bar{X}_{\sigma\nu\lambda}(\omega, \omega_1, \omega_2)$ includes 
energy denominators of triple-resonance to the input frequencies
$\omega_1$, $\omega_2$, and the observed frequency $\omega$ written as

\begin{eqnarray}
\bar{X}_{\sigma\nu\lambda}(\omega,\omega_1,\omega_2) 
&&
=
\frac{1}
{(\hbar\omega_1-\hbar\omega-i\gamma_\sigma)
(E_{\sigma\nu}-\hbar\omega-i\Gamma_\sigma)}\cdot 
\nonumber
\\
&&\hspace{-30mm}
\{\frac{1}{E_{\sigma\lambda}-\hbar\omega_2-i\Gamma_\sigma}+\frac{1}
{-E_{\sigma\lambda}+\hbar(\omega_1+\omega_2-\omega)-i\Gamma_\sigma}\} \nonumber 
\\
&&
\hspace{-30mm}
+\frac{1}
{(E_{\sigma\nu}-E_{\sigma\lambda}-\hbar\omega+\hbar\omega_2-i\Gamma_\sigma)
(E_{\sigma\nu}-\hbar\omega-i\Gamma_\sigma)} \cdot
\nonumber
\\
&&\hspace{-30mm}
\{
\frac{1}{-E_{\sigma\lambda}+\hbar(\omega_1+\omega_2-\omega)-i\Gamma_\sigma}
+\frac{1}{E_{\sigma\nu}-\hbar\omega_1-i\Gamma_\sigma}
\},
\label{X3}
\end{eqnarray}
where $\gamma_\sigma$ is a nonradiative population decay constant.
It should be noted that Eq. (\ref{X3}) has the same form as the case of the DFWM in Eq. (19) in Ref. \cite{Kinoshita}.
Generally, the OKR and the DFWM are different nonlinear processes; thus, the combination of incident frequencies changes for each process.
For example, assuming that the two incident lights are continuous waves with the pump frequency $\omega_1$ and the probe frequency $\omega_2$, the observed OKR frequency is $\omega=\omega_2$.
On the other hand, the observed DFWM frequency is $\omega=2\omega_1-\omega_2$ or $\omega=2\omega_2-\omega_1$.
Therefore, if we use the combination of $\omega_1$ and $\omega_2$ as the observed frequency instead of $\omega$, the expression for the denominator of the triple-resonance term has a different form for each process.
However, by utilizing the same $\omega$ for each process, we find that Eq. (\ref{X3}) is the same as that in the case of the DFWM, although the numerator differs for each process as shown in Eq. (\ref{U}) in this manuscript and Eq. (18) in Ref. \cite{Kinoshita}.
In addition, the generated nonlinear signal includes every combination of Fourier components in the pump and probe pulses. 
We integrate over these components by the numerical method.

As conditions of the input lights, we assume Gaussian pulses whose integrated intensity of the pump (probe) light is fixed to 3.0 $\mu$J/cm$^2$ (0.3 nJ/cm$^2$), and the center energies are both 3.378 eV. 
Here, we consider two temperature regions CT and RT, and set the corresponding non-radiative damping parameters as $\Gamma_\sigma=\gamma_\sigma=2$ and 30 meV.
In addition, we used the material parameters of A and B excitons in bulk ZnO \cite{Lagois} as listed in Table \ref{parameters}, where $m_0$ is the static electron mass.

\begin{table}[h]
\caption{Parameters of bulk ZnO \cite{Lagois}}
\begin{tabular}{l@{\hspace{25mm}}l}
\hline \hline
\hspace{10mm}A & \hspace{10mm}B \\ \hline
$M_A=0.87m_0$ & $M_B=0.87m_0$ \\
$E_A=3.3758$ eV & $E_B=3.3810$ eV \\
$E_{L1}=3.3776$ eV & $E_{L2}=3.3912$ eV \\
$\Delta_{LT1}=1.8$ meV & $\Delta_{LT2}=10.2$ meV \\
\hline
\label{parameters}
\end{tabular}
\end{table}

\begin{figure}[h]
\begin{center}
\includegraphics[width=85mm]{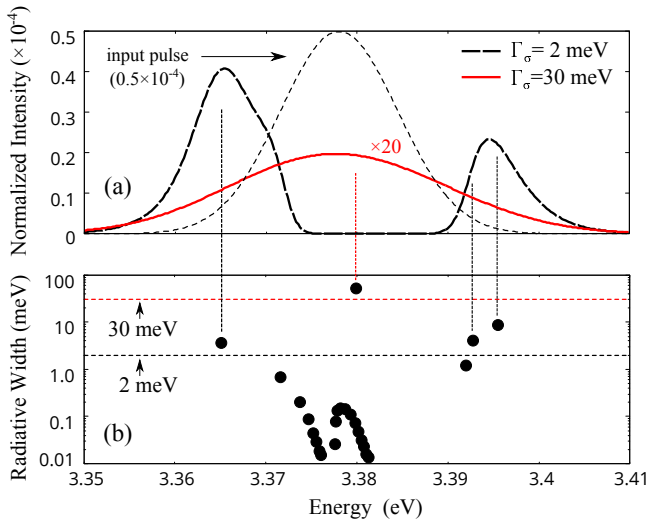}
\caption{(a) $\Gamma_\sigma$ dependence of calculated optical Kerr spectra $I_y(\omega)$ normalized by the peak intensity of the input probe pulse (the dotted line indicates spectrum of input 120-fs probe pulse), and (b) eigenmodes of exciton--radiation-coupled system (radiative width vs. eigenenergy) for a film thickness of 291 nm.}
\label{spectra}
\end{center}
\end{figure}

Figure $\ref{spectra}$ shows (a) the $\Gamma_\sigma$ dependence of the optical Kerr spectra $I_y (\omega)$ of a ZnO thin film normalized by the peak intensity of the input probe pulse $I_y(\omega)$, and (b) eigenmodes of the exciton--radiation-coupled system (radiative width vs. eigenenergy) for a film thickness of 291 nm.
The 120-fs input pulses can cover a certain range of spectral width and partially excite the lower (${\rm Re}[\hbar\omega_\xi]<E_A$), middle ($E_A\le{\rm Re}[\hbar\omega_\xi]\le E_B$), and upper ($E_B<{\rm Re}[\hbar\omega_\xi]$) branches of exciton--radiation-coupled modes at once.
The energy and spectral width of the signals obviously reflect the eigenenergy and radiative width. 
In the case where $\Gamma_\sigma=2$ meV (CT region), the eigenmodes with a radiative width larger than 2 meV dominantly appear; thus, large splitting between the lower and upper peaks is obvious in the spectrum.
On the other hand, the eigenmodes with a radiative width smaller than 2 meV are less reflected in the spectrum because of the damping effect.
Here, each of the two peaks at the CT should not be attributed solely to the A or B exciton according to their energy position.
These two peaks should be assigned to respective mixed modes containing both A and B excitons owing to the radiative coupling \cite{Kinoshita}.

With an increase in thermal damping $\Gamma_\sigma$, these two peaks disappear, and only one mode with a very broad spectral width appears (red line).
The radiative width of this mode is larger than 50 meV; thus, a large nonlinear signal can survive even at $\Gamma_\sigma=30$ meV (RT region) because the response is faster than the nonradiative damping under this condition.

In the aforementioned discussions, the 120-fs input pulse may not be very effective for maximizing the nonlinear intensity, because it cannot cover the entire radiative width of the fastest mode.  
Additionally, there are dependences of the input pulse width and the film thickness on the output nonlinear intensities.
In the next section, we investigate how the nonlinear intensity can be maximized by changing these parameters.

\section{Nonlinear efficiency \label{4}}
To evaluate the intensity of the OKR signals, we defined the nonlinear efficiency $\eta$ as the ratio of the integrated intensity of the input probe light to that of the output Y-polarized one, as follows:

\begin{eqnarray}
\eta=\frac{\int I_y(\omega)d\omega}{\int I_{probe}(\omega)d\omega}.
\end{eqnarray}

\begin{figure}[b]
\begin{center}
\includegraphics[width=85mm]{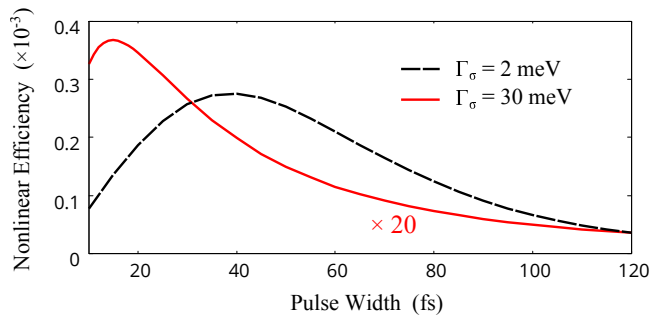}
\caption{Pulse-width dependence of nonlinear efficiency $\eta$ for a film thickness of 291 nm.}
\label{pulse_width}
\end{center}
\end{figure}

Figure \ref{pulse_width} shows the pulse-width dependence of the nonlinear efficiency $\eta$ for a film thickness of $291$ nm.
The value of $\eta$ is maximized when the input pulse effectively covers the peak structures as shown in Fig. \ref{spectra} (a). 
In the case where $\Gamma_\sigma=2$ meV (CT region), the upper and lower modes are relatively dominant compared with the fastest decay mode.
Therefore, nearly 40-fs pulses are optimal.
Pulses that are too short are not effective because of the loss of energy.

On the other hand, in the case where $\Gamma_\sigma=30$ meV (RT region), the fastest decay mode with a radiative width over 50 meV (radiative decay time reaching several femto seconds) dominantly survives in the optical responses.
In this situation, a shorter input with a nearly 15-fs pulse width is more effective for covering the very broad radiative width.
The nonlinear efficiency is enhanced by a factor of ～10 compared with that of the 120-fs pulse excitation. 
Significantly, the high damping does not greatly reduce the nonlinear efficiency for nearly 15-fs input pulses, although the damping is 15 times greater than that at CT. 
This is because the radiative width of the fastest mode is far larger than $\Gamma_\sigma$ and less affected by the thermal damping effects.

\begin{figure}[h]
\begin{center}
\includegraphics[width=85mm]{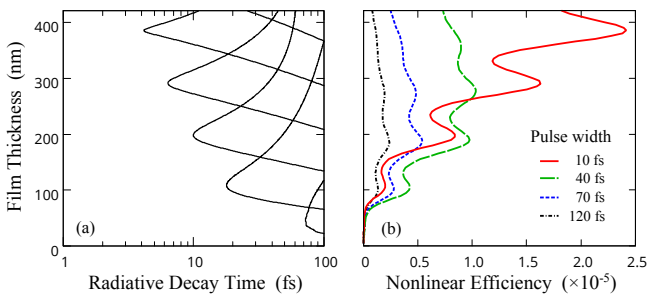}
\caption{Film-thickness dependence of (a) the radiative decay time $\tau_{\xi}$ of exciton--radiation-coupled modes, and (b) the nonlinear efficiency $\eta$ with different pulse widths at $\Gamma_\sigma=30$ meV.}
\label{integral}
\end{center}
\end{figure}

According to these results, we expect compatibility between fast and strong nonlinear responses caused by short-pulse excitation in the RT region.
We demonstrate this by comparing the radiative decay time $\tau_\xi$ and the nonlinear efficiency $\eta$.
Figure \ref{integral} shows the film-thickness dependences of (a) $\tau_{\xi}$ and (b) $\eta$ with different pulse widths at $\Gamma_\sigma=30$ meV.
These figures reveal two particularly noteworthy points.
(1) The optimal pulse width depends on the film thickness.
This is because the radiative width of the exciton--radiation-coupled mode is determined according to the interaction volume (film thickness) between the exciton and radiation field.
Therefore, when the radiative decay time becomes far shorter than the input pulse width with an increase in the film thickness, the nonlinear efficiency decreases.
(2) In the case of the 10-fs pulse, similarly to the tendency of the radiative decay times, the nonlinear efficiency increases with the film thickness.
In particular, the local minimal values of the radiative decay time and maximal values of the nonlinear efficiency occur around the same thickness, indicating that the fast radiative decay of excitons becomes compatible with the sufficient nonlinear responses if an appropriate film thickness and input pulse width are selected.

For a thickness region less than 80 nm in length, the nonlinear efficiency is weak.
However, around this thickness, where the speed of radiative-decay exceeds that of dephasing, the growth of the nonlinear efficiency becomes rapid even in the case of $\Gamma_\sigma=30$ meV. 
For ZnO, this occurs in a relatively thin region because particular exciton--radiation-coupled modes exhibit a larger radiative width owing to the radiative coupling of the A and B excitons.

\section{Conclusion \label{5}}
With the A and B excitons in ZnO, we theoretically investigated the radiative decay times of the exciton--radiation-coupled modes and their nonlinear optical responses by pulse excitation, focusing on the optical Kerr effect.
Because of the large radiative coupling of excitons, particular modes exhibit large radiative widths over 50 meV (short radiative decay times reaching several femto seconds), which can exceed the typical thermal damping at RT.
This is why such modes survive in coherent nonlinear optical signals such as the OKR in the case of a large damping parameter in the RT region.
Then, we demonstrated the pulse-width dependence of the nonlinear efficiency (defined as the ratio of the integrated intensity of the input light to that of the output light), which is expected to increase when the input pulse effectively covers the broad radiative widths of the exciton--radiation-coupled modes.
We discovered that the optimal short pulse enhanced the nonlinear efficiency with sufficient values even at RT, compared with those in the CT region.
Furthermore, the film-thickness dependence of the radiative decay times and nonlinear efficiency clearly indicate the possibility of compatibility between the fast and strong nonlinear responses obtained by choosing an appropriate film thickness and input pulse width.
The presented results draw a contrast to conventional observations of the optical responses of excitons, where a coherent nonlinear response is considered to be never prominent at RT, because the thermal dephasing is far faster than the typical radiative decay of excitons.  

The authors thank Professor M. Nakayama, Professor M. Ashida, Professor M. Ichimiya, and Professor N. Yokoshi for their fruitful discussions. 
This work was partially supported by JSPS KAKENHI Grant No. JP16H06504 in Scientific Research on Innovative Areas: ``Nano-Material Optical Manipulation."


\end{document}